\documentclass[preprint,prb,showpacs,preprintnumbers,amsmath,amssymb,superscriptaddress]{revtex4}
\usepackage{dcolumn}
\usepackage{bbm}
\usepackage{graphicx,epsfig}
\usepackage{bbm}
\begin{document}
\title{\bf{Effect of dimerization on dynamics of spin-charge separation in Pariser-Parr-Pople model: A 
time-dependent density matrix renormalization group study}}

\author{Tirthankar Dutta}
\affiliation{Condensed Matter Theory Unit, Jawaharlal Centre for Advanced Scientific Research, Jakkur Post, Bangalore
560064, India.}
\affiliation{Solid State and Structural Chemistry Unit, Indian Institute of
Science, Bangalore 560012, India}
\email{tirthankar@sscu.iisc.ernet.in}
\author{S. Ramasesha}
\email{ramasesh@sscu.iisc.ernet.in}
\affiliation{Solid State and Structural Chemistry Unit, Indian Institute of
Science, Bangalore 560012, India}

\begin{abstract}
We investigate the effect of static electron-phonon coupling, on real-time dynamics of spin and charge 
transport in $\pi$-conjugated polyene chains. The polyene chain is modeled by the Pariser-Parr-Pople 
Hamiltonian with dimerized nearest-neighbor parameter $t_{0}(1+\delta)$ for short bonds and $t_{0}(1-\delta)$
for long bonds, and long-range electron-electron interactions. We follow the time evolution of the spin and 
charge using time-dependent density matrix renormalization group technique, when a hole is injected at one end of the chain in its ground 
state. We find that spin and charge dynamics followed through spin and charge velocities, depend both on 
chain length and extent of dimerization, $\delta$. Analysis of the results requires focusing on physical 
quantities such as average spin and charge polarizations, particularly in the large dimerization limit. In 
the dimerization range 0.0 $\le$ $\delta$ $\le$ 0.15, spin-charge dynamics is found to have a well defined 
behavior, with spin-charge separation (measured as the ratio of charge velocity to spin velocity) as well 
as, the total amount of charge and spin transported in a given time, along the chain, decreasing as 
dimerization increases. However, in the range 0.3 $\le$ $\delta$ $\le$ 0.5, it is observed that the dynamics 
of spin and charge transport becomes complicated. It is observed that for large $\delta$ values, spin-charge 
separation is suppressed and the injected hole fails to travel the entire length of the chain.
\end{abstract}


\maketitle

\section{INTRODUCTION}
With vast advancements in technology, low-dimensional $\pi$-conjugated organic systems in recent times, have 
found use in single-molecule electronic and spintronic devices.\cite{ratner,reed,ritcher,expt1,expt2,expt3,expt4} 
Until now, these materials have been used in devices such as organic light-emitting diodes (OLEDS) and organic thin-film 
transistors.\cite{torsichap7,katzchap7,markschap7,nitzanchap7} The $\pi$-conjugated organic materials form an 
interesting class of strongly correlated systems in which there exists long-range electron correlations. 
Therefore, the low-energy physics of these systems is different from low-dimensional strongly correlated 
materials described by the Hubbard model. In order to propose and design $\pi$-conjugated organic systems as 
components of electronic (spintronic) circuits, a proper theoretical understanding of the mechanism of charge
and spin transport in these systems is thus necessary. Theoretical understanding of transport in 
many-particle systems with strong correlations requires appropriate techniques and formulations, mainly 
because transport is essentially an {\it out-of-equilibrium} phenomena. The advent of time-dependent density matrix renormalization group 
(td-DMRG) technique has vastly helped in addressing this issue. 
\cite{lxwchap7,white1chap7,daleychap7,white2chap7,scholRMPchap7} Even so, spin and charge transport in 
$\pi$-conjugated systems have not been addressed until recently using the td-DMRG method due to the fact that, 
most of the existing td-DMRG algorithms are chiefly structured to handle short-range electron-electron 
interactions. Furthermore, those few that are capable of handling long-range interactions suffer from 
drawbacks of large computational resources for their study. The {\it double time window targeting} (DTWT) 
technique proposed by us \cite{duttachap7}, has been employed to address the issue of 
dynamics of spin and charge transport in $\pi$-conjugated systems.

The issue of coupling of the electronic and lattice degrees of freedom on the stability of the metallic
state in one-dimensional conductors, has been of interest for a long time. 
\cite{frohlich,peirels,ooshika,higgins} One-dimensional systems such as 
polyenes, typified by trans-polyacetylene ($t$-PA), belong to the class of $\pi$-conjugated molecular 
materials having linear (chain) topology. These materials are the simplest $\pi$-conjugated systems that 
have been studied extensively both experimentally and theoretically. \cite{soosrama,mazumdardixit}  In these 
systems, electronic structure is strongly affected by electron-phonon interactions leading to dimerization, 
which is stabilized and enhanced by electron-electron correlations. As a result, these systems have 
dimerized ground state and their low-energy excitations are gapped, the extent of which depends both on the 
degree of dimerization and electron-electron interactions. Thus, it is to be expected that electron-phonon 
coupling will also influence the {\it out-of-equilibrium} dynamics of spin-charge separation in particular, 
and transport in general. Hence, an understanding of the role of dimerization on the dynamics of spin and 
charge transport in these systems is of considerable interest. Like most studies, we have also treated the 
electron-phonon coupling in the adiabatic limit (or Born-Oppenheimer approximation), i.e., the phonon degrees
of freedom are considered as slow, classical variables. In this limit, the sole effect of electron-lattice 
coupling in polyenes manifests as bond alternation, the extent of which is dictated by the  alternation (or 
dimerization) parameter $\delta$. A non-zero value of $\delta$ implies that the nearest-neighbor hopping 
integrals alternate as $t_{0}(1 \pm \delta)$. The object of this study is to investigate the effect of
this bond alternation on the dynamics of spin and charge transport in one-dimensional systems typified by 
polyenes.
The organization of the paper is as follows: In Sec. II, we discuss in detail, the model Hamiltonian and 
computational strategy used. Section III presents the results of our study along with discussions. In 
Sec. IV, we present our conclusions.
 
\section{MODEL AND COMPUTATIONAL METHODOLOGY}
The Pariser-Parr-Pople (PPP) model \cite{ppp1chap7,ppp2chap7} is appropriate for investigating the effect of 
dimerization on the dynamics of spin and charge transport in polyenes. In the second quantized representation,
the PPP Hamiltonian reads as
\begin{equation}
\begin{split}
\hat{H}_{\text{PPP}} &= \sum_{i=1}^{L-1} \sum_{\sigma} t_{0}[1-(-1)^{i} \delta] (\hat{c}^{\dagger}_{i,\sigma}\hat{c}_{i+1,\sigma} + \hat{c}^{\dagger}_{i+1,\sigma}\hat{c}_{i,\sigma}) \\
               &+ \sum_{i=1}^{L} \frac{U_{i}}{2}\hat{n}_{i}(\hat{n}_{i}-1) + \sum_{j>i} V_{ij} (\hat{n}_{i}-z_{i})(\hat{n}_{j}-z_{j}).  
\end{split}
\end{equation}
{\noindent
Here, $L$ denotes the number of carbon atoms in the polyene chain with open boundary condition, 
$\hat{c}^{\dagger}_{i,\sigma}$ ($\hat{c}_{i,\sigma}$) creates (annihilates) an electron with spin orientation $\sigma$ 
on the $i^{\text{th}}$ carbon atom, $t_{0}$ is the average transfer integral without dimerization and, 
0 $\le$ $\delta$ $\le$ 1 is the bond alternation or dimerization parameter. The strength of on-site Coulomb 
repulsion between two electrons of opposite spins on site $i$ is $U_{i}$ and $\hat{n}_{i}$ is the electron 
number operator for the same site. The term $V_{ij}$ represents the inter-site Coulomb repulsion between sites 
($i, j$), with $z_{i}$ being the on-site chemical potential of the $i^{\text{th}}$ carbon atom. Polyenes 
being homogeneous $sp^{2}$ carbon systems, $U_{i}$ = $U$ at all sites, and to maintain charge neutrality when
a site is singly occupied, we also set $z_{i}$ = 1 for all $i$. The inter-site interaction between electrons
on sites $i$ and $j$, $V_{ij}$, is interpolated between $U$ for $r_{ij}$ = 0 and $\frac{e^2}{r_{ij}}$ for 
$r_{ij} ~\rightarrow ~\infty$ by Ohno interpolation\cite{ohno} given by,}
\begin{equation}
V_{ij} ~=~ 14.397(1.6348+r_{ij}^{2})^{-1/2}.
\end{equation}
{\noindent
where $V_{ij}$s are in eV and $r_{ij}$s are in \AA. Although there is an algebraic fall off in the inter-site potential 
$V_{ij}$, all (equilibrium) properties such as optical gap, two photon gap and spin gap can be extrapolated to the 
thermodynamic limit. This is because the transfer part of the Hamiltonian is short-ranged and the effective correlation 
strength, $V_{\text{eff}}$ = ($U$ - $V_{12}$) is smaller than the band-width of the one-particle spectrum. Aside from this, the 
$\pi$-coherence length (which is the length of the $\pi$-system beyond which intensive properties such as excitation 
gaps, saturate)in such system is only about 20 sites long. Hence we expect that (non-equilibrium) properties such as 
spin and charge velocities, discussed in the subsequent section, also approach the thermodynamic limit.
In this study we deal with polyene chains of 20, 30 and 40 carbon atoms, $\delta$ is set to 0.0, 0.05, 0.07, 0.15, 0.3 
and 0.5, and the rest of the parameters assume standard values for the PPP model for $t$-PA and polyenes 
\cite{rama1chap7,rama2chap7,rama3chap7,rama4chap7}: $t_{0}$ = $-$2.4 eV, $U$ = 11.26 eV, and $2\pi/3$ bond angle between 
successive bonds. The PPP Hamiltonian possesses charge-conjugation and inversion symmetries, and also conserves total 
spin. Dimerization affects the transfer term and the distance-dependent electron-electron repulsions ($V_{ij}$) only. It 
does not influence the on-site Coulomb repulsion ($U$) between electrons.} 

As $\delta$ increases from 0.0 to 0.5, hopping integrals for partial double bonds get enhanced from $-$2.4  
to $-$3.6 eV, while those for partial single bonds reduce from $-$2.4 to $-$1.2 eV. As a result, the $1$-norm of 
the Hamiltonian matrix increases, implying that $E_{c}$ = max$[|E_{\text{max}}|,|E_{\text{min}}|]$ also 
increases. The dimensionless time-step $\alpha$ = $E_{c}$ $\Delta t$ of a numerical scheme for solving the 
time-dependent Schr\"odinger equation, defines its stability region, with $\Delta t$ being the time-step of 
evolution \cite{iitakachap7}. The value of $\alpha$ is constant for a given ordinary differential equation 
(ODE) solver, and hence as $E_{c}$ increases, $\Delta t$ has to be decreased. Thus, with increase in 
dimerization, the time-step for propagating the Schr\"odinger equation forward in time, has to be reduced for
numerical stability. This is the scenario with commonly used ODE solvers such as the Runge-Kutta (RK) schemes, 
the Crank-Nicholson (CN) method, and the multi-step differencing (MSD) techniques. Since the DTWT technique 
\cite{duttachap7} uses the MSD2 scheme \cite{iitakachap7,msd2} for updating the Hilbert basis and the 
fourth-order RK technique for time evolution, with increase in dimerization the DTWT procedure becomes 
computationally time consuming. Hence, we modified the DTWT algorithm by replacing these two time evolution methods with 
the Chebyshev-polynomial-based expansion of the time evolution operator, $\hat{U}(\Delta t)$ = 
$\exp(-i\hat{H}\Delta t)$.\cite{kosloffchap7,lubichchap7} The Chebyshev-polynomial-based scheme has the 
advantage that the expansion of $\hat{U}(\Delta t)$ can be evaluated up to machine accuracy and is free from 
any time-step constraint. 

The Chebyshev-polynomial-based time evolution involves propagating the state $\mid \psi(t) \rangle$ by time-step 
$\Delta t$ by approximating the discrete time evolution operator in terms of
Chebyshev polynomials as 
\begin{equation}
\begin{split}
e^{-i\hat{H}\Delta t}   &\approx \sum_{m=0}^{P} a_{m} T_{m}(\hat{\mathbbm{H}}),
\end{split}
\end{equation}
where, $T_{m}(\hat{\mathbbm{H}})$ is the $m^{\text{th}}$ Chebyshev polynomial of the first kind, $\hat{\mathbbm{H}}$ 
represents the scaled Hamiltonian with eigenvalues ranging from $[-1.0, 1.0]$, and the coefficients $a_{m}$ are given by
\begin{equation}
a_{m} = (2-\delta_{m0})e^{-i\Delta t \gamma}(-i)^{m}J_{m}(\Delta t \beta),
\end{equation}
where, $\gamma$ = $(E_{\text{max}}+E_{\text{min}})/2$ and $\beta$ = 
$(E_{\text{max}}-E_{\text{min}})/2$; $E_{\text{max}}$, $E_{\text{min}}$ are the maximum and minimum 
eigenvalues of the PPP Hamiltonian, and $J_{m}$ is the $m^{\text{th}}$ order Bessel function of the first 
kind. The necessity of scaling $\hat{H}_{\text{PPP}}$ to $\hat{\mathbbm{H}}$ arises from the argument domain 
of the Chebyshev polynomials 
of the first kind $T_{m}(x)$; $x$ $\in$ $[-1 , 1]$. The Chebyshev polynomials can be generated using the 
following recursion relation \cite{KPMrmp,prb2011},
\begin{equation}
T_{m+1}(\hat{\mathbbm{H}}) \mid \psi(t) \rangle = \biggl[2\hat{\mathbbm{H}}T_{m}(\hat{\mathbbm{H}}) - T_{m-1}(\hat{\mathbbm{H}}) \biggr]\mid \psi(t) \rangle,
\end{equation}            ̃
{\noindent
with the initial conditions, $T_{0}(\hat{\mathbbm{H}})$ $\mid \psi \rangle$ = $\mid \psi \rangle$, and 
$T_{1}(\hat{\mathbbm{H}})$ $\mid \psi \rangle$ = $\hat{\mathbbm{H}}$ $\mid \psi \rangle$. However, since the 
coefficients $a_{m}$ are known in advance [Eq. (4)], instead of using this forward recursion scheme, we use 
the ``reverse'' recursion algorithm proposed by Clenshaw \cite{clenshaw,fox}, which is more stable. The 
Clenshaw recursion requires $P$ sparse-matrix vector multiplications (SMVMs) of the Hamiltonian $\mathbbm{H}$ 
with the state vector $\mid \psi(t) \rangle$. When $P$ $>$ $\frac{1}{2}$ $\Delta t$ 
$(E_{\text{max}}-E_{\text{min}})$, the error decays almost exponentially \cite{talchap7}. In case of 
increase of $1$-norm of the Hamiltonian matrix, one needs to either retain a higher $P$ or reduce the 
magnitude of time step $\Delta t$. We increase the value of $P$ with enhancement in dimerization.} 

The basic DTWT algorithm remains unchanged when ODE solvers such as the fourth-order RK and MSD2 schemes are 
replaced by the Chebyshev-polynomial-based expansion of the time-evolution operator. However, for a given number of 
retained density matrix eigenvectors (DMEVs) ($m$) and a given time step ($\Delta t$), results of the Chebyshev-polynomial-based DTWT td-DMRG algorithm are more 
accurate than the ODE-based version of the algorithm. This is because the ODE solvers have truncation errors 
associated with them; the truncation errors associated with the MSD2 and the fourth-order RK 
procedures are $O[(\Delta t)^3]$ and $O[(\Delta t)^5]$, respectively.\cite{duttachap7,iitakachap7,msd2} The Chebyshev-polynomial-based expansion of the time evolution operator, on the other hand, is free from such truncation errors and 
hence, can be evaluated up to machine accuracy by keeping $P$ (number of SMVMs) $>$ 
$\frac{1}{2} \Delta t (E_{max}-E_{min})$.\cite{talchap7} However, the increase in accuracy is significant only at 
longer times; in the initial stages of time evolution, the accuracy is similar for the ODE and the Chebyshev-polynomial-based DTWT algorithms. 

The number of SMVMs associated with the basic MSD2 and fourth-order RK steps are $1$ and $4$, respectively. 
For a system with $N$ sites, in the finite-system DMRG algorithm, for a single full sweep we have $4(N/2-2)$ basic steps.
Thus, total SMVMs for a single full sweep are $4(N/2-2)$ $+$ $4$. Since each single time window $\Delta t$ is subdivided
into $p$ time slices of width $\Delta \tau$, for propagating the wave packet by a single window, the total number of SMVMs
is $8p(N/2-2)$ $+$ $4p$. The first term corresponds to the MSD2 scheme for a full finite sweep of the $N$ site system
over two time windows and the second corrresponds to propagation of the wave packet by a single time window. The total 
SMVMs per single time window are $4p(N-3)$ \cite{duttachap7}. In case of the Chebyshev-polynomial method, the total 
number of SMVMs per single time window $\Delta t$ is $4P(N/2-2)$ or $2P(N-4)$, $P$ is dependent on $\Delta t$, and 
usually $P$ $>$ $p$. However, $\Delta \tau$ in ODE methods should be very small ($\sim$ $10^{-4}$ fs) when the 1-norm of 
the Hamiltonian matrix is large, for reasonable accuracy, while in the Chebyshev-based method, $\Delta \tau$ $\sim$ 
$10^{-1}$ fs can be employed for similar accuracy. Thus, the number of time steps in ODE based methods ($p_{\text{ODE}}$)
is larger than in Chebyshev polynomial methods ($p_{\text{CP}}$). This leads to $p_{\text{CP}} \times P$ $\sim$ 
$p_{\text{ODE}}$ and the two approaches require approximately the same computational time.

In order to investigate the effect of dimerization on the dynamics of spin and charge transport, an up spin 
electron is annihilated from the first carbon site of the polyene chains of $L$ sites in its half-filled
ground state $\mid \phi^{0}_{gs} \rangle$, thereby leading to an initial wave packet $\mid \psi(0) \rangle$
\begin{equation}
\mid \psi(0) \rangle = \hat{c}_{1,\uparrow} \mid \phi_{gs}^{0} \rangle.
\end{equation}
This wave packet is propagated in time by solving the time-dependent Schr\"odinger equation numerically, 
using the Chebyshev polynomial-based expansion of $\hat{U}(\Delta t)$. Using the time evolved wave packets 
$\mid \psi(t) \rangle$, site charge density $[\langle \hat{n}_{i}(t) \rangle]$, and site spin density 
$[\langle \hat{s}_{i}^{z}(t) \rangle]$, at time $t$ are computed as, 
\begin{align}
\langle \hat{n}_{i}(t) \rangle &= \langle \psi(t) \mid (\hat{n}_{i,\sigma} + \hat{n}_{i,-\sigma}) \mid \psi(t) \rangle, \\
\langle \hat{s}^{z}_{i}(t) \rangle &= \frac{1}{2} \langle \psi(t) \mid (\hat{n}_{i,\sigma} - \hat{n}_{i,-\sigma}) \mid \psi(t) \rangle.
\end{align}
{\noindent
These quantities give the dynamics of the injected hole in terms of its spin and charge degrees of 
freedom. The real-time dynamics of the initial wave packet is studied using a modified DTWT scheme, wherein 
adaptation of the Hilbert space as well as time evolution, are performed with the Chebyshev polynomial-based 
expansion of time evolution operator. The other parameters used in our study are: DMEVs retained, 
$m$ = 300; time step for evolution $\Delta \tau$ = 0.066 fs;
total evolution time $T$ = 33.0 fs. However, for the purpose  of this study we focus only on the initial 
15.0 fs, as this is adequate for our purpose. In the modified DTWT procedure, weights of the reduced density 
matrices of all the time-dependent wave packets are kept the same, unlike in 
the originally formulated scheme,\cite{duttachap7} since 
we are using a large step size of 0.066 fs. Although time-dependent charge and spin densities are computed at
all the sites, the quantities $\langle \hat{n}_{1}(t) \rangle$ and $\langle \hat{n}_{L}(t) \rangle$, and 
$\langle \hat{s}^{z}_{1}(t) \rangle$ and $\langle \hat{s}^{z}_{L}(t) \rangle$ are sufficient for 
investigating the effect of dimerization on the dynamics of the hole in terms of its spin and charge degrees 
of freedom; $1$ and $L$ correspond to the first and last sites of the $\pi$-conjugated chain with $L$ sites.} 

\section{RESULTS AND DISCUSSION}
The temporal variation in site charge density and site spin density for the terminal sites of a PPP chain 
of $40$ sites are shown in Figs. 1 and 2, for various dimerization strengths $\delta$. The first 
significant dip in $\langle \hat{n}_{L}(t) \rangle$ and $\langle \hat{s}^{z}_{L}(t) \rangle$ correspond to 
the times $\tau^{h}_{L}$ and $\tau^{s}_{L}$, taken by the charge and spin of a hole injected at the first
site, to reach the end of the chain, respectively. The second significant dip in the time evolution plots of 
$\langle \hat{n}_{1}(t) \rangle$ and $\langle \hat{s}^{z}_{1}(t) \rangle$ occurs when the charge and spin
degrees of freedom of the hole, get reflected back from the chain end and reach the injection site. These
two dips correspond to times $\tau^{h}_{2L}$ and $\tau^{s}_{2L}$, respectively. This is supported by the 
fact that $\tau^{h}_{2L}$ and $\tau^{s}_{2L}$ are very close to twice the values of $\tau^{h}_{L}$ and 
$\tau^{s}_{L}$, respectively. From these times, the charge velocity ($\vartheta^{h}_{L}$) and spin 
velocity ($\vartheta^{s}_{L}$) are calculated in a straight forward way: 
$\vartheta^{h/s}_{L}$ = $\biggl(L/\tau^{h/s}_{L}\biggr)$. From Figs. 1 and 2 it is observed that locating 
$\tau^{h/s}_{L}$ is easy for dimerizations $\delta$ up to 0.15. For dimerizations $\delta$ = 0.3 and 0.5, 
it is however very hard to locate these minima unambiguously. This may be due to the fact that the weak 
bonds, when very weak (large $\delta$), do not easily transmit the charge or spin resulting in interference 
of the wave traveling forward with the reflected wave. This effect is seen more in the longer chains as the 
charge or spin needs to travel through many weak bonds.

\begin{figure}[!tbp]
\begin{center}
\includegraphics[scale=0.5]{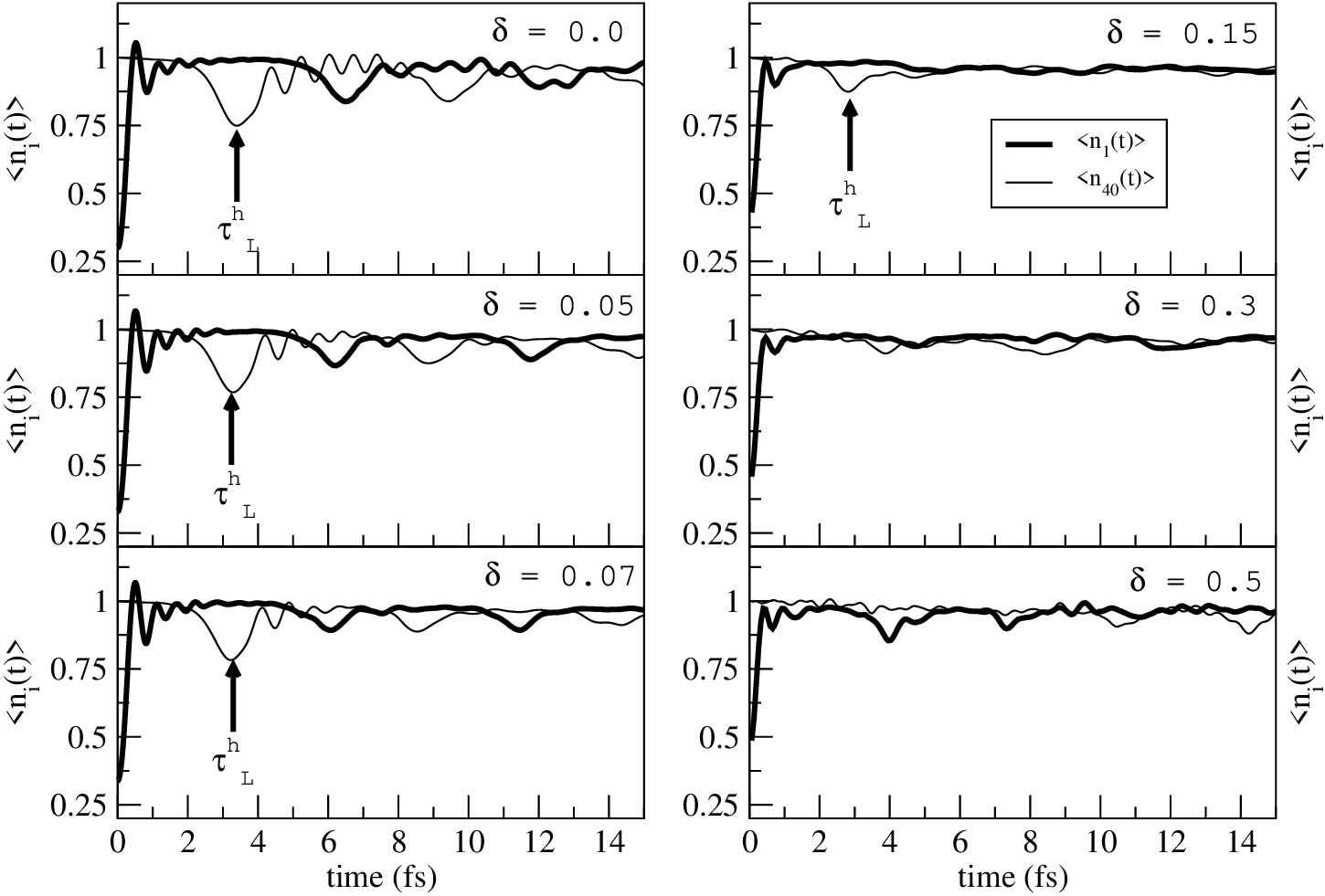}  
\end{center}
\caption{Temporal variation in charge densities, $\langle n_{1}(t) \rangle$ and $\langle n_{40}(t) \rangle$, 
at sites $1$ and $40$ in the PPP model, for different dimerizations, $\delta$, for a chain of 40 sites. In 
each box, $\langle n_{1}(t) \rangle$ is shown by thick black curve and $\langle n_{40}(t) \rangle$, by thin 
black curve. For 0.0 $\le$ $\delta$ $\le$ 0.15, $\tau^{h}_{L}$ is indicated by arrow.} 
\end{figure}

\begin{figure}[!tbp]
\begin{center}
\includegraphics[scale=0.5]{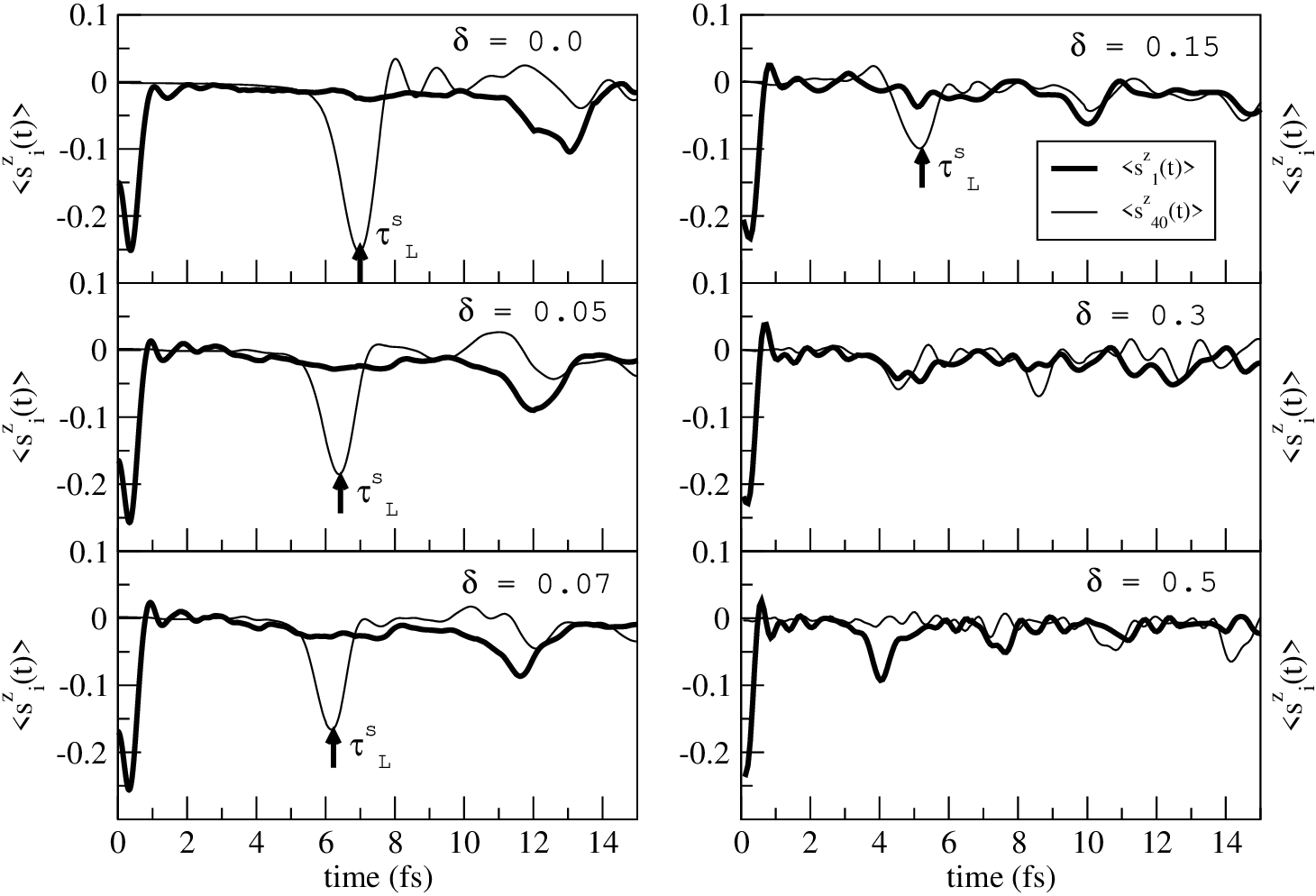}  
\caption{Temporal variation in spin densities, $\langle s^{z}_{1}(t) \rangle$ and 
$\langle s^{z}_{40}(t) \rangle$, at sites $1$ and $40$, in the PPP model, for different dimerizations, 
$\delta$, for a 40-site chain. In each box, $\langle s^{z}_{1}(t) \rangle$ is shown by thick black curve and 
$\langle s^{z}_{40}(t) \rangle$, by thin black curve. For 0.0 $\le$ $\delta$ $\le$ 0.15, $\tau^{s}_{L}$ is 
indicated by arrow.} 
\end{center}
\end{figure}

Normally, charge and spin velocities are discussed in the bulk limit. From our earlier studies on the one-dimensional 
Hubbard model,\cite{duttachap3} we have noticed that it is not possible to relate the analytically obtained charge and 
spin velocities $\vartheta_{\rho/\sigma}$, with $\vartheta^{h/s}_{L/2L}$ obtained for finite small systems. This is 
because, the analytic velocities are obtained from the {\it exact} excitation spectrum as group velocities in the limit 
of momentum $\vec{q}$ $\rightarrow$ 0,\cite{coll} while in our case $\vartheta^{h/s}_{L}$ pertain to a state which is 
not an eigenstate of the Hamiltonian governing the dynamics of the system. Furthermore, dimerization introduces a gap in 
the excitation spectrum and there is no low energy theory which completely separates the spin and charge degrees of 
freedom. This is also true for the one-dimensional PPP model [Eq. 1] for which however, no analytical expressions for 
the spin and charge velocities are available in the literature. Hence, unlike $\vartheta_{\rho/\sigma}$, 
$\vartheta^{h/s}_{L}$ depends on system size. This dependence is more pronounced in the H\"uckel and Hubbard models, 
compared to the PPP model, due to the absence of long-range interactions.\cite{duttachap7} The ratio of the velocities, 
$\frac{\vartheta^{h}_{L}}{\vartheta^{s}_{L}}$, due to finite size effects is system size dependent. We need to go to 
longer chains (longer $L$) for this ratio to be size independent. For the PPP model however, this dependence is weak 
(see Fig. 3). 

In Table I we present the times taken by the charge and spin of the hole to propagate from one end of the 
chain to the other end, as well as, the charge and spin velocities, and their ratios. The times taken by the 
spin and charge to reach the end of the chain approximately scale with chain length. However, the velocities 
of both charge and spin increase slightly with increasing chain length, which is due to weak finite-size 
effects. In Table I, we also present the spin and charge velocities extrapolated to infinite system size for
different dimerization strengths ($\delta$ = 0.0, 0.05, 0.07 and 0.15), as well as, the ratio 
$\vartheta^{h}_{\infty}/\vartheta^{s}_{\infty}$. We see that the ratio $\vartheta^{h}_{\infty}/\vartheta^{s}_{\infty}$ 
decreases with an increase in $\delta$. 
In Fig. 3 is shown the dependence of the ratio of charge to spin velocities as a function of 
dimerization, for the chain lengths studied. We find from the plot that the ratio decreases as $\delta$ 
increases. Indeed, within the error of resolution of $\tau^{h/s}_{L}$, even for $\delta$ = 0.3 and 0.5, it 
appears that this trend continues and the ratio of the velocities approaches the non-interacting value of 
$1.0$. This can be contrasted with the push-pull polyene systems studied by us,\cite{duttaarxiv} where the
ratio of the spin and charge velocities is independent of the strength of the push-pull groups.
The above feature, namely, dependence of $\vartheta^{h}_{L}/\vartheta^{s}_{L}$ on $\delta$, cannot be attributed to 
change in interactions brought about by geometry changes as a consequence of increased dimerization. This is because 
these changes are small and it has been shown in earlier studies \cite{ramaalbert} that the contributions to the energy 
gaps between states due to changes in interaction parameters, $V_{ij}$, accompanying small bond length changes (due to
change in $\delta$), is rather small. Therefore it appears that in dimerized models, the change in transfer 
integrals has a stronger role to play than the long-range interactions. 

\begin{figure}[!tbp]
\begin{center}
\includegraphics[scale=0.5]{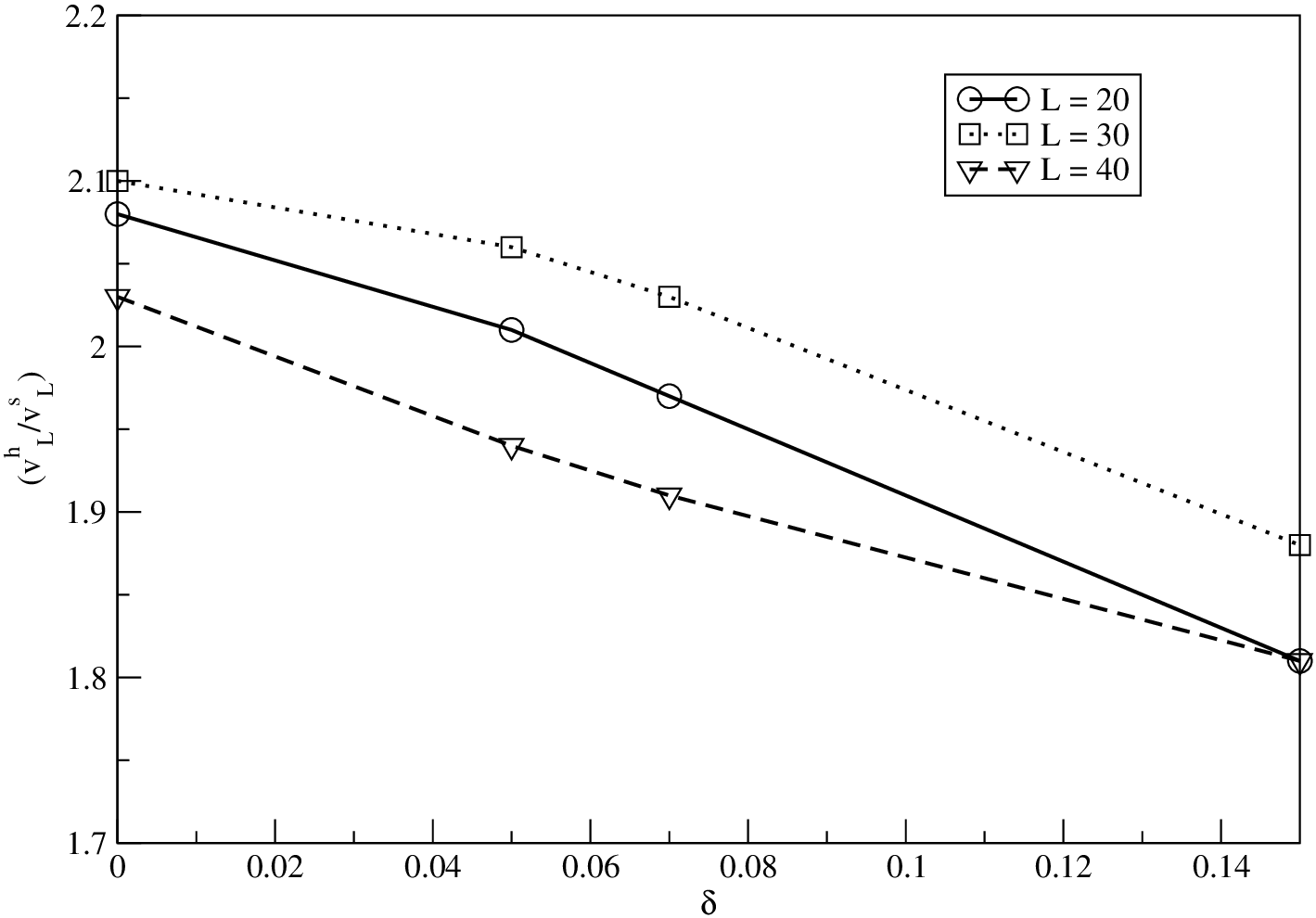} 
\caption{Variation in the ratio of charge and spin velocities ($\vartheta^{h}_{L}/\vartheta^{s}_{L}$) with 
dimerization 0.0 $\le$ $\delta$ $\le$ 0.15, in the PPP model, for different chain lengths.} 
\end{center}
\end{figure}
\vspace{1.5cm}
\begin{table}[!tbp]
\caption{Variation in the times $\tau^{h}_{L}$, $\tau^{s}_{L}$ (fs), and velocities $\vartheta^{h}_{L}$, 
$\vartheta^{s}_{L}$ (\AA/fs), and ratio of velocity of charge to velocity of spin 
$\vartheta^{h}_{L}/\vartheta^{s}_{L}$, in the PPP model with dimerization, for different chain lengths,
$L$ = $(N-1)r_{0}$ - $\delta$; $N$ is the number of sites and $r_{0}$ = 1.397 \AA is the $C=C$ bond length in a regular polyene chain.}
\begin{center}
\begin{tabular}{|c||c|c|c|c|c|c|}
\hline
$N$ & $\delta$ & $\tau^{h}_{L}$ & $\tau^{s}_{L}$ & $\vartheta^{h}_{L}$ & $\vartheta^{s}_{L}$ & $(\vartheta^{h}_{L}/\vartheta^{s}_{L})$ \\
\hline\hline
20  & 0.0  & 1.81 & 3.77 & 14.66 &  7.04 & 2.08 \\
    & 0.05 & 1.76 & 3.54 & 15.05 &  7.48 & 2.01 \\
    & 0.07 & 1.75 & 3.45 & 15.13 &  7.67 & 1.97 \\ 
    & 0.15 & 1.72 & 3.12 & 15.34 &  8.46 & 1.81 \\ 
\hline
30  & 0.0  & 2.59 & 5.44 & 15.64 &  7.45 & 2.10 \\
    & 0.05 & 2.49 & 5.15 & 16.25 &  7.87 & 2.06 \\
    & 0.07 & 2.45 & 4.98 & 16.51 &  8.12 & 2.03 \\ 
    & 0.15 & 2.24 & 4.22 & 18.02 &  9.56 & 1.88 \\ 
\hline
40  & 0.0  & 3.44 & 6.97 & 15.84 &  7.82 & 2.03 \\
    & 0.05 & 3.29 & 6.39 & 16.54 &  8.52 & 1.94 \\
    & 0.07 & 3.23 & 6.17 & 16.85 &  8.82 & 1.91 \\ 
    & 0.15 & 2.84 & 5.15 & 19.13 & 10.55 & 1.81  \\ 
\hline
$\infty$  & 0.0  & - & - & 17.14 &  8.52 & 2.01 \\
          & 0.05 & - & - & 18.15 &  9.35 & 1.94 \\
          & 0.07 & - & - & 18.71 &  9.75 & 1.92 \\ 
          & 0.15 & - & - & 22.99 & 12.43 & 1.85  \\ 
\hline
\end{tabular}
\end{center}
\end{table} 

In order to investigate in detail the issue of the ratio $\frac{\vartheta^{h}_{L}}{\vartheta^{s}_{L}}$ 
tending to the non-interacting value of 1.0 as $\delta$ $\rightarrow$ 1.0, we have focused on the time 
dependence of the charge (spin) polarization defined as, $\vec{P}_{c(s)}(t)$. These are calculated from 
the normalized site charge (spin) density $\rho^{c(s)}_{i}(t)$, with background correction. Employing these
renormalized quantities suppresses the large fluctuations seen in $\langle n_{i}(t) \rangle$ and
$\langle s^{z}_{i}(t) \rangle$ allowing one to focus on the essential behavior of spin and charge transport.
The quantities $\rho^{c}_{i}(t)$ and $\rho^{s}_{i}(t)$ are defined as  
\begin{equation}
\rho^{c}_{i}(t) = \frac{\langle n_{i}(t) \rangle - C}{\sum_{i}(\langle n_{i}(0) \rangle - C)};~~\rho^{s}_{i}(t) = \frac{\langle s^{z}_{i}(t) \rangle - S}{\sum_{i}(\langle s^{z}_{i}(0) \rangle - S)}, 
\end{equation}
where $C$ and $S$ are the average background charge and spin densities respectively, and are given by, 
\begin{equation}
C = \frac{1}{N}\sum_{j}\langle \phi^{0}_{gs} \mid n_{j} \mid \phi^{0}_{gs} \rangle;~~S = \frac{1}{N}\sum_{j}\langle \phi^{0}_{gs} \mid s^{z}_{j} \mid \phi^{0}_{gs} \rangle.
\end{equation}
For a half-filled ground state belonging to the $S^{z}_{tot}$ = 0 subspace, $C$ = 1.0 and $S$ = 0.0. 
Using these observables, we define time-dependent charge and spin polarizations as, $\vec{P}_{c(s)}(t)$ = 
$\sum_{j} \rho^{c(s)}_{j}(t) \vec{r}_{j}$. The time-dependence of the charge and spin polarizations qualitatively 
reflect the center-of-mass movement of the charge and spin peaks.\cite{kanhare} However, these polarizations are gauge 
dependent as the system is having a net charge. Hence we have computed these by choosing as origin, the central site of 
a $N$ site chain ($N$ = 20, 30, 40). 

\begin{figure}[!tbp]
\begin{center}
\includegraphics[scale=0.5]{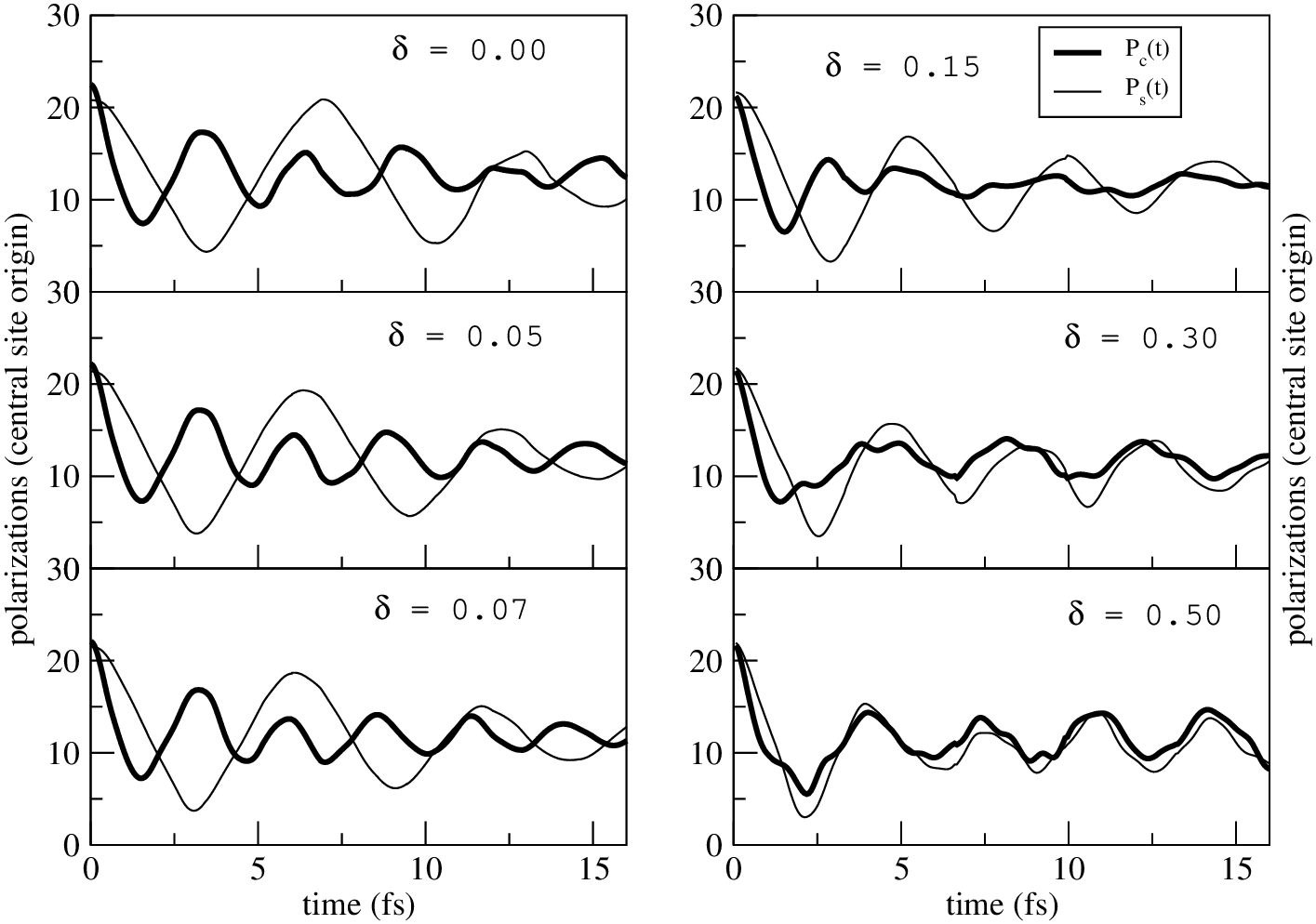}  
\caption{Temporal variation in $\vec{P}_{c}(t)$ (dark curves) and $\vec{P}_{s}(t)$ (light curves) 
for a $40$ site PPP chain, for different dimerizations, $\delta$.} 
\end{center}
\end{figure}

In Fig. 4, we have plotted the variations in charge and spin polarizations with time, for a polyene chain
of $40$ sites. Motion of the charge (spin) across the two ends of the chain leads to oscillations in the 
temporal variation of $\vec{P}_{c(s)}(t)$. The larger the amplitude of the oscillations, the higher is 
the probability of charge (spin) traveling between the chain ends. It is observed from both the curves that, 
with an increase in dimerization, the oscillation amplitude decreases. This indicates that electron-phonon 
coupling reduces the probability of both the charge and spin degrees of freedom to travel between the two 
ends of the chain. This observation indeed supports our earlier observation that it is difficult to locate 
$\tau^{h/s}_{L}$ for $\delta$ = 0.3 and 0.5 merely because, weak bonds fail to transmit the charge or spin 
easily. In other words, the charge and spin for large values of $\delta$, might not even reach the opposite 
end of the chain, especially for long chains. We also notice that an increase in dimerization suppresses 
spin-charge separation, which is manifested by the fact that $\vec{P}_{c}(t)$ and $\vec{P}_{s}(t)$ closely 
follow each other. Thus, indeed the ratio of $\frac{\vartheta^{h}_{L}}{\vartheta^{s}_{L}}$ $\rightarrow$ 1.0
as $\delta$ $\rightarrow$ 1.0. In other words, with an increase in electron-phonon coupling, spin-charge 
separation is suppressed in the PPP model.

However, the question about fate of the charge and spin of the injected hole for large values of $\delta$, 
still remains. In order to investigate this issue, we have computed
\begin{eqnarray}
\frac{dN_{\text{RL}}(t)}{dt} = \frac{N_{\text{R}}(t)-N_{\text{L}}(t)}{\Delta t}, \\
\frac{dS^{z}_{\text{RL}}(t)}{dt} = \frac{S^{z}_{\text{R}}(t)-S^{z}_{\text{L}}(t)}{\Delta t}.
\end{eqnarray}
Here, $N_{\text{R/L}}(t)$ = $\sum_{j \in R/L} \langle n_{j}(t) \rangle$ and $S^{z}_{\text{R/L}}(t)$ = 
$\sum_{j \in R/L} \langle s^{z}_{j}(t) \rangle$; $R$ $\in$ $[N/2+1,N]$ and $L$ $\in$ $[1,N/2]$. 
These quantities provide information about the amount of charge and spin of the hole that is transported 
from the left-half to the right-half of the chain. In the initial state of the system, both the charge and 
spin of the hole reside solely in the left-half of the system. Hence, $\frac{dN_{\text{RL}}(t)}{dt}$ and 
$\frac{dS^{z}_{\text{RL}}(t)}{dt}$ at time $t$ = 0 have the values 1.0 and 0.5, respectively. With time, as 
the charge and spin travel from the first to the last site of the chain, these quantities change sign and 
depending on the strength of electron correlations, attain values close to $-$1.0 and $-$0.5. When the charge and
spin travel back from the last to the first site, these quantities again change sign and reach values close 
to $+$1.0 and $+$0.5. This thus indicates that, $\frac{dN_{\text{RL}}(t)}{dt}$ and 
$\frac{dS^{z}_{\text{RL}}(t)}{dt}$ oscillate between $\pm 1.0$ and $\pm 0.5$, respectively. The amplitude of 
oscillation of these quantities with time gives a measure of the amount of charge and spin that has been 
transported from the left-half to the right-half of the system. If under some circumstance the charge (spin) 
fails to travel between the ends of the chain, time evolution of these two quantities does not not show any 
sign change.   

\begin{figure}[!tbp]
\begin{center}
\includegraphics[scale=0.5]{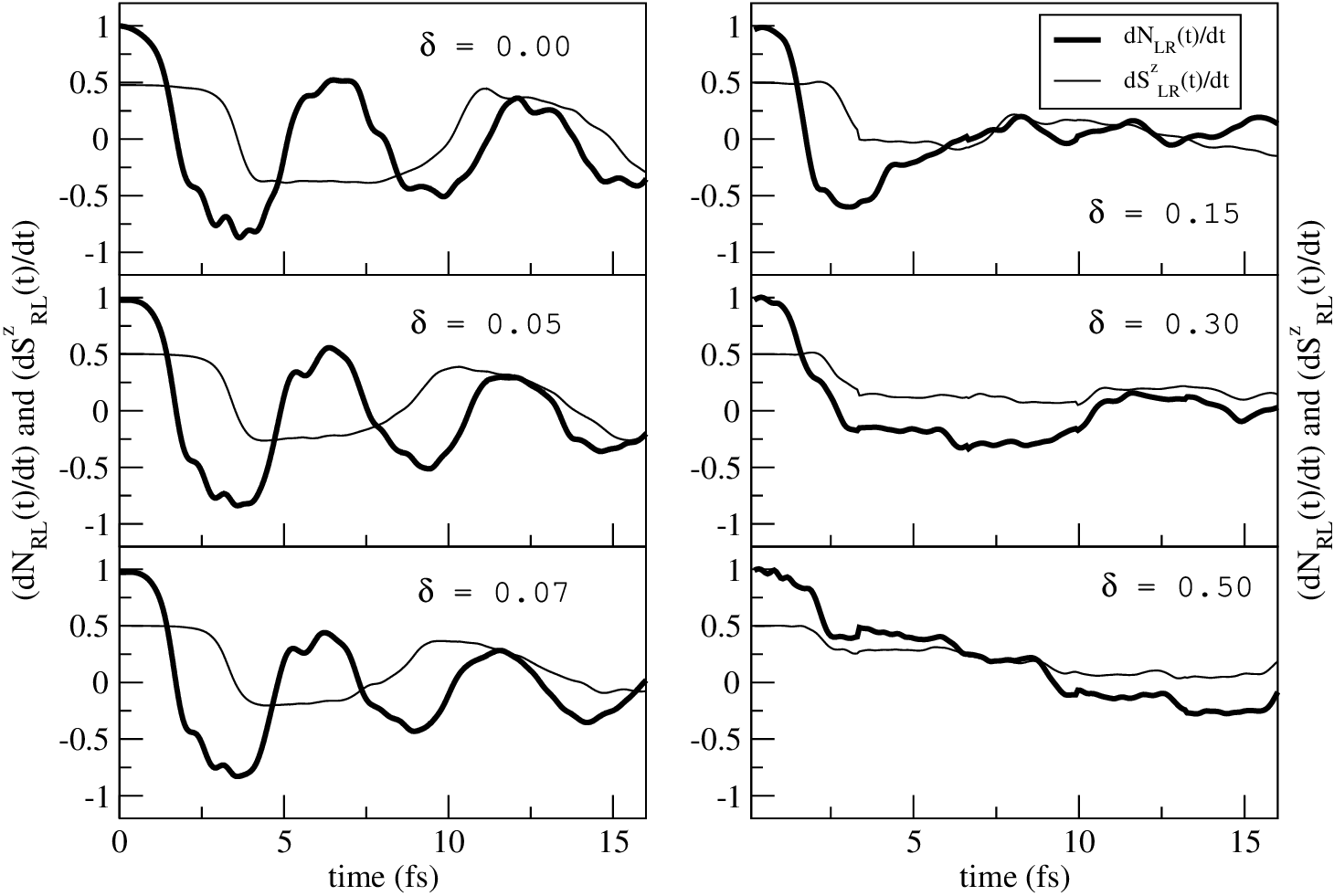} 
\caption{Variation in $\frac{dN_{\text{RL}}(t)}{dt}$ (dark curves) and $\frac{dS^{z}_{\text{RL}}(t)}{dt}$ 
(light curves) with time, for a $40$ site PPP chain, for different dimerization strengths: $\delta$ = 0.0, 
0.05, 0.07, 0.15, 0.3, and 0.5.} 
\end{center}
\end{figure}

Figure 5 depicts the time evolution of $\frac{dN_{\text{RL}}(t)}{dt}$ and $\frac{dS^{z}_{\text{RL}}(t)}{dt}$ 
for a polyene chain of $40$ sites, for different dimerization strengths. It is observed that these quantities
indeed oscillate between $\pm 1.0$ and $\pm 0.5$, although the amplitude of oscillations decreases with time, 
as $\delta$ increases. For small dimerizations ($\delta$ = 0.05 and 0.07), it is observed that the temporal 
variations in $\frac{dN_{\text{RL}}(t)}{dt}$ and $\frac{dS^{z}_{\text{RL}}(t)}{dt}$ are similar to those in 
the undimerized case, and the number of times $\frac{dN_{\text{RL}}(t)}{dt}$ and 
$\frac{dS^{z}_{\text{RL}}(t)}{dt}$ changes sign is the same for all the three cases. However, the number of 
``sign changes'' in the time propagation of $\frac{dN_{\text{RL}}(t)}{dt}$ is more than that of 
$\frac{dS^{z}_{\text{RL}}(t)}{dt}$, indicating that the charge degree of freedom moves faster compared to the
spin degree of freedom which is merely a manifestation of spin-charge separation. Thus, for small values of 
electron-phonon coupling the motion of the charge and spin of the hole do not get hindered due to 
interferences. 

However for $\delta$  = 0.15, 0.3 and 0.5, the time evolution profiles of $\frac{dN_{\text{RL}}(t)}{dt}$ and 
$\frac{dS^{z}_{\text{RL}}(t)}{dt}$ are completely different from those obtained with small values of 
$\delta$. It is observed that the amplitude of oscillations diminishes with time indicating that the amount 
of charge and spin transported across the chain decreases as electron-phonon coupling increases. This 
supports our observation that an increase in $\delta$ leads to a decrease in the height of the minima, in the 
time evolution profiles of $\langle n_{L}(t) \rangle$ and $\langle s^{z}_{L}(t) \rangle$. Also, the number of 
times that $\frac{dN_{\text{RL}}(t)}{dt}$ and $\frac{dS^{z}_{\text{RL}}(t)}{dt}$ changes sign decreases, which
reflects that strong electron-phonon coupling significantly reduces the probability of to-and-fro motion of 
the charge and spin along the chain. For $\delta$ = 0.3 and 0.5, however, $\frac{dS^{z}_{\text{RL}}(t)}{dt}$ 
does not change sign within the time scale of our studies. This depicts that, for these values of $\delta$, 
only the charge degree of the hole can travel across the length of the chain, albeit slowly and in a very small
amount. The spin of the hole fails to reach the end of the chain and ``gets trapped'' within the left-half
of the chain. This is the reason why we find that locating $\tau^{h/s}_{L}$ for large values of $\delta$ 
is difficult.

Bosonization studies on one-dimensional spinful Hubbard and Luttinger liquid models, at half-filling 
\cite{gia,mocanu,schuster}, indicate that $\delta$ $\ne$ $0$ introduces a backward-scattering operator into the 
bosonized Hamiltonain, which is associated with momentum transfer $q$ $\sim$ $4k_{F}$. Because of this umklapp 
scattering term, as dimerization increases, propagation of the charge and spin degrees of freedom of the injected hole 
between the ends of the chain become hindered, and for large values of $\delta$ (0.3 and 0.5), fail to reach the chain 
end ($L$). Furthermore, periodic perturbation of the lattice such as Peierls distortion, which introduces a gap in the
excitation spectrum, has been shown to couple the charge and spin sectors of the one-dimensional (bosonized) 
Hubbard Hamiltonian.\cite{schuster} These two results, obtained with respect to spinful fermion models with short-range 
electron-electron interactions, seem to support both our observations in the context of the PPP model, namely, an increase 
in dimerization suppresses spin-charge decoupling, and inhibits the oscillatory to-and-fro motion of the charge and spin 
of the injected hole. 

\section{CONCLUSION}
To conclude, we have shown that dimerization modifies the dynamics of spin and charge transport in the PPP
model. For small dimerizations the system shows a smooth decrease in the ratio of charge to spin velocities,
although both the velocities decrease slightly. For large values of $\delta$, the situation is different.
It is difficult to obtain the charge and spin velocities as strong interference effects smear out their 
motion. However, a more careful analysis based on net charge transport across the chain shows that both
the charge and spin stay almost localized for the duration of our time evolution. Thus, although dimerization
of polyenes predominantly changes the one-electron part of the Hamiltonian, it strongly affects spin-charge
separation. Large values of dimerization reduce the amount of charge and spin transported across the 
system as well as, suppress spin-charge separation, as evidenced by the decrease in 
$\vartheta^{h}_{L}/\vartheta^{s}_{L}$ ratio and by closely following time evolution profiles of spin and
charge polarizations. 

~\\
\begin{center}
{\bf ACKNOWLEDGMENTS}
\end{center}
~\\
This work was supported by DST India and the Swedish Research Link Program under the Swedish Research 
Council. T.D. wishes to acknowledge S. Sahoo for stimulating and useful discussions.

\end{document}